\documentclass[conference]{IEEEtran}
\IEEEoverridecommandlockouts

\usepackage{cite}
\usepackage{amsmath,amssymb,amsfonts}
\usepackage{algorithmic}
\usepackage{graphicx}
\usepackage{textcomp}
\usepackage{xcolor}
\def\BibTeX{{\rm B\kern-.05em{\sc i\kern-.025em b}\kern-.08em
    T\kern-.1667em\lower.7ex\hbox{E}\kern-.125emX}}
\begin{document}

\title{Decentralized Multi-agent Reinforcement Learning for Resilient Critical Infrastructures}



\author{
\IEEEauthorblockN{
\begin{tabular}{cc}
Minghui Ding\IEEEauthorrefmark{1}
&
Evangelos Pournaras\IEEEauthorrefmark{1}\IEEEauthorrefmark{2}
\\
kknx3385@leeds.ac.uk
&
E.Pournaras@leeds.ac.uk
\end{tabular}
}

\IEEEauthorblockA{
\IEEEauthorrefmark{1}
\textit{School of Computer Science, University of Leeds}, Leeds, UK
}

\IEEEauthorblockA{
\IEEEauthorrefmark{2}
\textit{School of Energy Systems, LUT University}, Lappeenranta, Finland
}
}

\maketitle

\begin{abstract}
Critical infrastructures are increasingly distributed, interdependent, and exposed to evolving disruptions, making resilience a central requirement for their operation and control. This paper argues that decentralized multi-agent reinforcement learning (MARL) should be understood not merely as a distributed alternative to centralized training with decentralized execution but as a paradigm structurally aligned with the requirements of resilient critical infrastructures. This perspective is grounded in an analysis of the properties of decentralized MARL and the requirements of critical infrastructures, including scalability to large numbers of agents, support for privacy and local autonomy, robustness to failures, and interaction-driven adaptation among interdependent components. However, structural alignment alone is insufficient for practical deployment. This paper identifies credit assignment and communication as two central conditions for its practical feasibility. Credit assignment determines whether local learning remains aligned with system-level objectives, while communication determines whether coordination can be learned and maintained under realistic operational constraints. Building on these challenges, this paper proposes a research agenda focused on structure-aware, causality-aware, and resilience-aware credit assignment; communication for both coordination and credit assignment; and safe, timely, and recoverable decentralized learning under deployment constraints. Overall, this paper reframes decentralized MARL as a promising but conditional foundation for resilient critical infrastructures.

\end{abstract}


\section{Introduction}
Critical infrastructures such as transportation systems~\cite{Chu19}, smart grids~\cite{wang2021multi}, water networks~\cite{negm2024deep}, and industrial control systems~\cite{stouffer2011guide} are increasingly integrating digital information systems with operational technologies and physical processes. Although such integration enables more responsive monitoring, control, and optimization, it also strengthens the coupling between digital systems and the physical processes they support, expands the attack space of operational systems, and complicates resilience management under failures and disruptions\cite{pournaras2020cascading, thapa2019measuring}. In such settings, infrastructures must not only operate efficiently under normal conditions, but also sustain or recover critical functions when parts of the system become degraded or unavailable. As a result, resilience becomes a central requirement for contemporary critical infrastructures.

Multi-agent reinforcement learning (MARL) provides a natural computational framework for studying these systems because it captures how multiple decision-making agents learn through interaction with their environment and with each other~\cite{panfili2018game,ghannad2021prioritizing}.
While traditional control methods remain fundamental in many infrastructure applications~\cite{ouyang2014review}, MARL is especially relevant when distributed components must make interdependent decisions and adapt to changing conditions. However, existing MARL research most commonly follows the paradigm of centralized training with decentralized execution (CTDE), in which agents act locally during execution but rely on centralized aggregation of data or global information during training~\cite{lowe2017multi}. Although this paradigm has been effective in many simulated settings, its underlying training assumptions are not always compatible with critical infrastructures, where global information is costly to obtain, communication is constrained, and operational data may not be freely aggregated across large-scale or cross-organizational systems~\cite{petrenj2021cross}. These limitations motivate a shift in attention from CTDE to decentralized MARL in the context of critical infrastructures. In decentralized MARL, agents learn and act on the basis of local observations and limited interactions, without relying on centralized aggregation of training data or access to global information. 

This paper argues that decentralized MARL is not merely a distributed alternative to CTDE, but a structurally appropriate paradigm for resilient critical infrastructures. This perspective is grounded in a systematic analysis of how the core properties of decentralized MARL align with the operational requirements of critical infrastructures. Specifically, decentralized MARL scales naturally to large numbers of agents by distributing learning and decision-making rather than relying on a single controller, accommodates privacy constraints and local autonomy through selective information exchange rather than centralized data aggregation, and maintains robustness under partial failures and degraded communication because neither training nor execution depends on complete system-wide information. Beyond these operational properties, this paper distinguishes decentralized MARL from decentralized federated learning, arguing that its interaction-driven learning structure is essential for infrastructure settings where resilience depends on mutual adaptation among interdependent components rather than isolated local optimization.

At the same time, structural alignment alone is not sufficient for practical deployment. This paper identifies credit assignment and communication as two central conditions for making decentralized MARL feasible in resilient critical infrastructures. Credit assignment concerns whether local agents can learn behaviors that support system-level resilience rather than only local performance, while communication concerns whether coordination can be learned and maintained under realistic constraints on bandwidth, latency, and reliability. Building on these challenges, the paper further develops a research agenda around three priorities: structure-aware, causality-aware, and resilience-aware credit assignment; communication for both coordination and credit assignment; and safe, timely, and recoverable decentralized learning under deployment constraints. Taken together, these considerations position decentralized MARL as a promising but conditional paradigm whose practical value depends on addressing these research priorities.

The remainder of this paper is organized as follows. Section \ref{sec:problem} outlines the problem setting and learning process for decentralized MARL. Section \ref{sec:correspondence} develops the structural correspondence argument in detail. Section \ref{sec:challenges} analyzes credit assignment and communication as two core challenges for practical deployment. Section \ref{sec:agenda} introduces a research agenda for decentralized MARL in resilient critical infrastructures. Section \ref{sec:conclusion} concludes the paper.

\section{Problem Setting and Learning Process for Decentralized MARL}
\label{sec:problem}

Distributed intelligence requires agents that can learn and coordinate without a central controller. In this section, we first outline a general problem model for decentralized MARL and then illustrate the corresponding learning process.

\subsection{Problem Setting}

We consider a group of autonomous agents operating in a shared environment and communicating through a peer-to-peer network, which may be fixed in some settings and time-varying in others. Each agent is equipped with local sensing and computation capabilities and makes decisions based solely on its own observations and limited information from others. Due to the absence of a centralized controller, agents learn to coordinate their behaviors in a decentralized manner to achieve a common system-level objective. 

A concrete real-world example of this setting is decentralized traffic signal control in an urban road network. Each intersection operates as an autonomous agent that observes only local traffic conditions, such as queue lengths, vehicle arrivals, waiting times, and possibly the current signal phases of neighboring intersections. In this example, the intersections are physically fixed, and the nominal communication graph may follow stable road adjacency. However, the effective communication graph among controllers is not necessarily fixed, since available links or selected communication neighbors may change over time. Each agent follows a control policy that maps its local observations to signal control actions, such as switching or extending traffic light phases. The agents aim to coordinate signal phases to reduce congestion and improve traffic flow across the network. In the absence of a centralized traffic controller, each intersection independently learns and updates its control policy using locally available feedback, such as local queue lengths, waiting times, and information exchanged with neighboring intersections. Although each agent optimizes a local objective, their interactions give rise to emergent coordination that improves the overall traffic efficiency, for example, by reducing congestion and travel time across the network.

This setting can be modeled as a decentralized MARL problem, in which multiple agents learn to make decisions in a shared, dynamic environment without a centralized controller and with only limited information exchange.

\subsection{Learning Process}

\begin{figure}[!t]
    \centering
    \includegraphics[width=\linewidth]{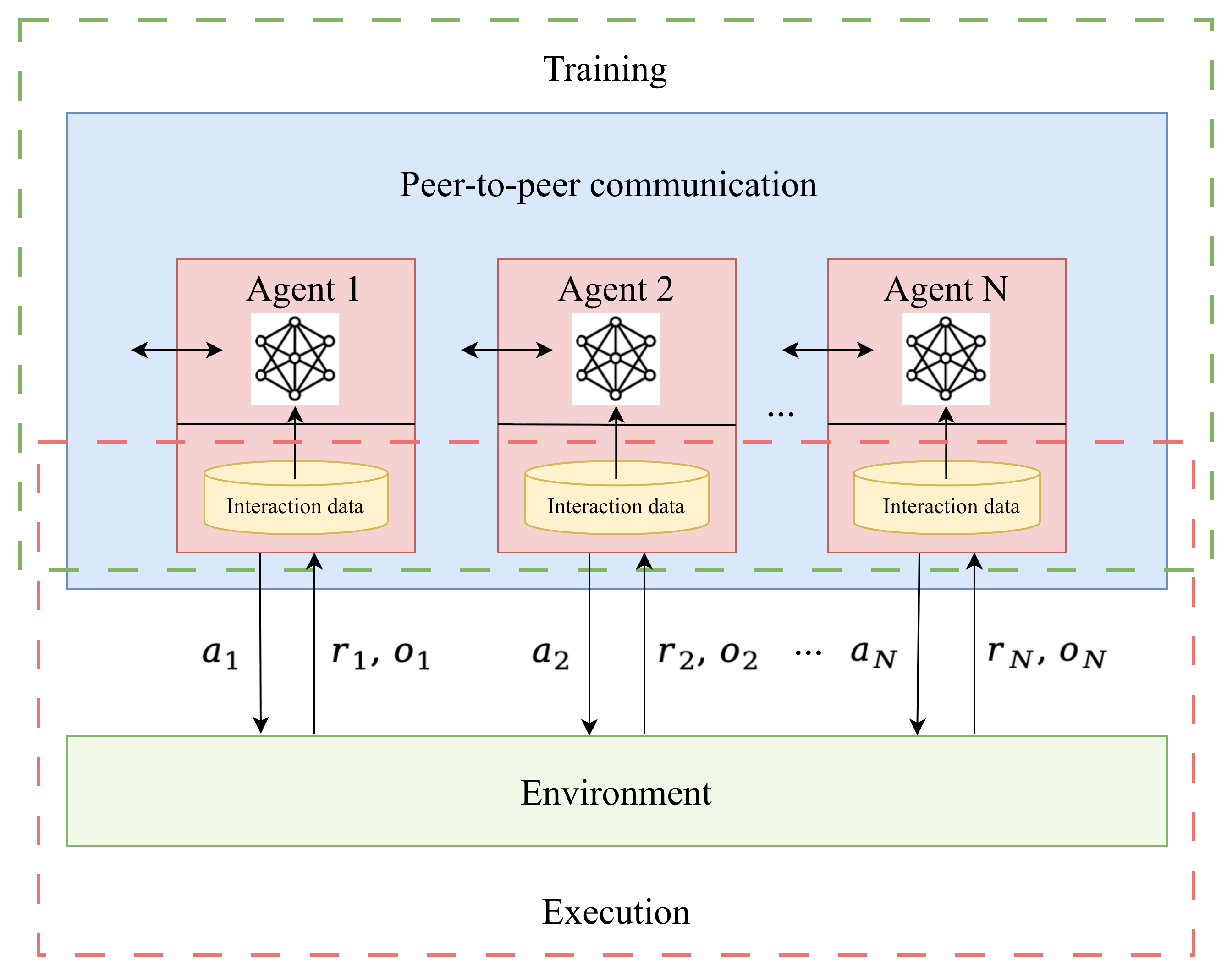}
    \caption{Decentralized MARL with peer-to-peer communication.
    During execution, each agent $i \in \{1, \dots, N\}$ selects an action $a_i$ based on its local observation $o_i$ and receives reward $r_i$ from the environment. During training, each agent $i$ updates their policies using their own interaction data and information exchanged with other agents through peer-to-peer communication.}
    \label{fig:networkedMARL}
\end{figure}

In decentralized MARL, both training and execution are fully distributed across agents. During training, agents update their policies using their own interaction data, comprising observations, actions, and rewards, augmented by information exchanged with other agents through peer-to-peer communication, without relying on any form of global guidance. During execution, each agent selects actions according to its learned decision rule (policy) based solely on its local observations and receives feedback in the form of local rewards from the environment. The goal of the agents is to improve the total reward accumulated over time. The architecture of the decentralized MARL system is illustrated in Fig.~\ref{fig:networkedMARL}.

\begin{figure}[!t]
  \centering
  \includegraphics[width=\linewidth]{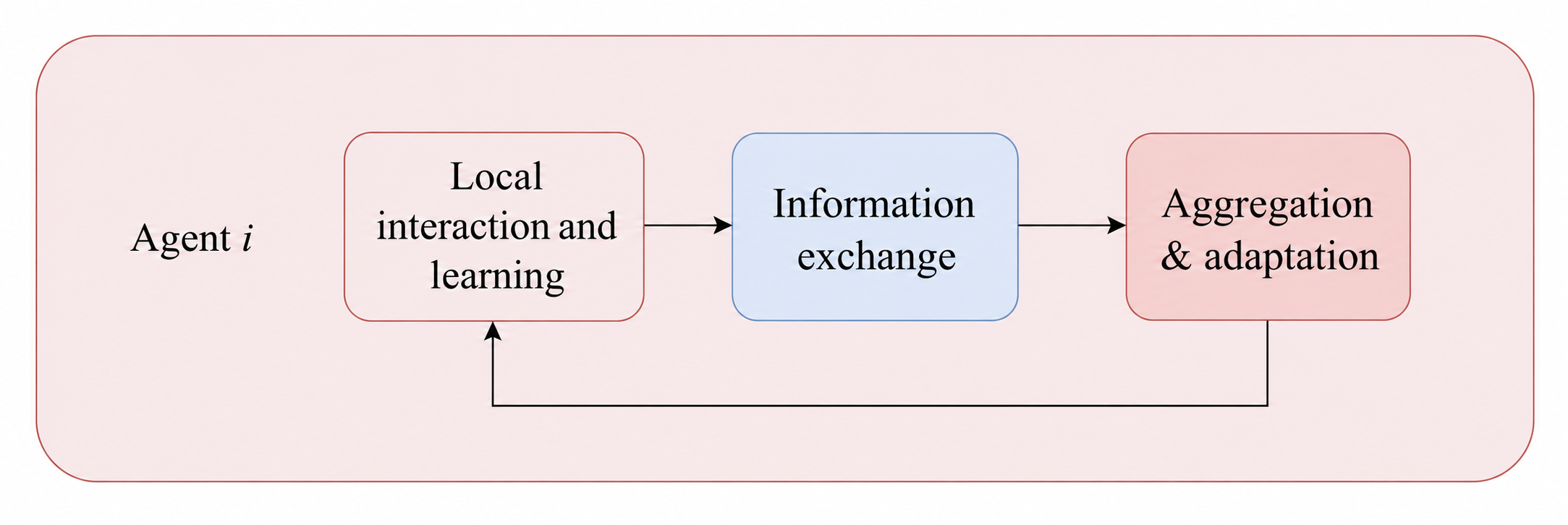}
  \caption{Iterative learning cycle of an individual agent in decentralized MARL. In each iteration, agent $i$ interacts with the environment and performs local learning using its own interaction data, exchanges learned information with neighboring agents, and aggregates the received information to adapt its local model before the next iteration.}
  \label{fig:framework}
\end{figure}

To describe the learning dynamics in more detail, we decompose the iterative learning cycle of each agent into three conceptual stages: local interaction and learning, information exchange, and aggregation and adaptation, as illustrated in Fig.~\ref{fig:framework}.

\textbf{Local Interaction and Learning}: 
During training, each agent learns a local decision-making rule that maps its own observations to actions, using feedback from the environment. Through repeated interactions, the agent collects sequences of local observations, selected actions, and received rewards, and uses this experience to improve its decision rule over time. In practice, this decision rule is typically represented using neural networks and updated with single-agent reinforcement learning methods~\cite{mnih2016asynchronous, schulman2017proximal,mnih2015human}. 

\textbf{Information Exchange}:
To support coordination, agents periodically exchange information with a limited set of neighboring agents through a peer-to-peer network. Rather than sharing raw data comprising observations, actions, and rewards, this communication focuses on higher-level summaries of what agents have learned, allowing cooperation without leaking privacy. Existing decentralized MARL methods provide different examples of parameter-sharing-based information exchange. In actor--critic settings, agents may exchange critic parameters~\cite{zhang2018fully}, actor parameters~\cite{ren2025communication}, or both~\cite{chen2022communication}, depending on the assumptions and communication budget. Similar ideas also apply beyond actor--critic methods, as value-based agents can coordinate through neighborhood-level exchange of neural-network parameters~\cite{malucelli2025neighbor}.

\textbf{Aggregation and Adaptation}:
Each agent incorporates information received from others into its own decision-making process and adapts its behavior accordingly.  This may involve reweighting alternative actions or adjusting how strongly the agent relies on its own past interactions versus information from peers. For example, each agent can take a weighted average of its own network parameters and those received from its neighbors~\cite{zhang2018fully}.
Through repeated interactions and information exchange, agents gradually adjust to one another, enabling coordinated behavior to emerge despite the information gap.

\section{Decentralized MARL and Critical Infrastructures}
\label{sec:correspondence}

Decentralized MARL is relevant for critical infrastructures not only because of its learning capability, but also because of the structural properties of its decentralized design. This section examines how these properties make it more suitable for large-scale, privacy-sensitive, and interdependent infrastructures, with particular attention to operational compatibility, robustness, and resilience.

\subsection{Operational Compatibility under Infrastructure Constraints}

Decentralized MARL is relevant for critical infrastructures because these systems are large, distributed, and often require timely local responses. As the number of components increases, collecting global information and computing system-wide decisions through a single controller can become expensive in terms of communication, computation, and latency. Decentralized MARL is well matched to such settings because learning and decision-making are distributed across agents rather than concentrated in a central controller~\cite{zhang2021multi}. Each agent processes local information and coordinates with a limited set of neighboring agents, allowing the system to scale without continuous centralized aggregation\cite{ma2024efficient}.

Another challenge is that several infrastructure systems cannot freely centralize data and control. Although this constraint is not the primary motivation in the traffic signal control example, it becomes more prominent in other infrastructure domains. In smart grids~\cite{pournaras2020holarchic, mcdaniel2009security}, for example, local entities such as microgrids and substations must optimize their operations while sharing only limited 
information, as detailed operational data may be sensitive or costly to aggregate centrally. Related work on decentralized collective learning has similarly shown how autonomous participants can coordinate energy-management decisions without centralizing private data\cite{pournaras2018decentralized}. Decentralized MARL accommodates similar constraints by enabling coordination through selective information exchange rather than requiring full central access to local data.

\subsection{Robustness under Disruptions}

Robustness is a central requirement in critical infrastructures, where the system must be able to sustain essential functions under disruptions such as component failures, communication degradation, and adversarial actions~\cite{rehak2018resilience, aldawsari2025optimization}. Centralized approaches can be fragile in such settings because they depend heavily on a central controller that can be a single point of failure. Once that central controller becomes unavailable or falls under adversarial control, the whole system may collapse.

Decentralized MARL is better suited to these conditions because agents are trained and deployed under partial observations and limited neighbor-to-neighbor information. Since their policies do not rely on complete system-wide state information or continuous centralized coordination, the system can degrade more gracefully under local disturbances, disrupted communication, and partial failures. In traffic signal control~\cite{Zhang23}, for example, the unavailability of one intersection or communication link does not necessarily imply the collapse of the full control system. Neighboring intersections may still continue operating on the basis of local traffic observations and whatever limited information remains available. In these situations, the value of decentralized MARL lies not only in distributed execution, but in allowing local components to keep learning and operating under incomplete information and partial disruption.

\subsection{Resilience through Interaction-Driven Adaptation}

While robustness concerns whether a system can continue operating during disruptions, resilience further requires that the system can recover, reorganize, and adapt after disruptions have altered operating conditions. This distinction is important for critical infrastructures because the effects of accidental failures and adversarial actions can propagate across tightly coupled cyber and physical components\cite{pournaras2020cascading, thapa2019measuring, pournaras2016self}. 

\begin{figure}[!t]
    \centering
    \includegraphics[width=\linewidth]{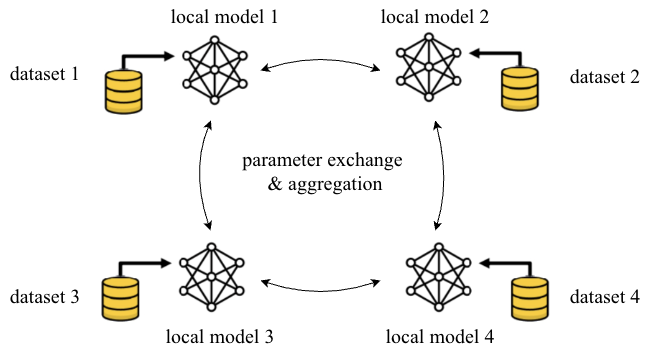}
    \caption{Decentralized federated learning over distributed local datasets. Each participant trains a local model on its own data and coordinates through parameter exchange and aggregation.}
    \label{fig:FL}
\end{figure}

Resilience in such settings depends not only on whether computation or decision-making is decentralized, but also on whether the learning process can capture how local actions influence the operating conditions of other components. This is where decentralized MARL differs from decentralized federated learning. In decentralized federated learning, distributed nodes typically optimize a shared and fixed learning objective over locally collected data~\cite{zhang2021surveyFL, hegedus2019gossip}, as shown in Fig.~\ref{fig:FL}. 
Although updates may be aggregated or exchanged, they do not directly alter the data generating processes of other participants. Consequently, any non-stationarity that arises in federated learning, such as data drift or system dynamics, is generally external to the learning process itself rather than driven by interactions among learners. In contrast, in decentralized MARL, each agent interacts with and alters the shared environment, thereby influencing the interaction data collected by other agents, as shown in Fig.~\ref{fig:networkedMARL}. Learning therefore involves mutual adaptation through the environment rather than isolated optimization, with the effective learning targets evolving as agents learn. 

This distinction is especially important for resilience in critical infrastructures, because disruptions in one component can reshape the operating conditions faced by others. In traffic signal control, for example, a disruption at one intersection can change upstream and downstream traffic flows, requiring neighboring controllers to adapt their signal policies in response. In such settings, the learning framework must also capture how interdependent components adapt to shared and evolving system conditions. Decentralized MARL therefore provides a more relevant framework for critical infrastructures in which resilience depends on interaction-driven adaptation. 

\section{Deployment Challenges of Decentralized MARL for Critical Infrastructures}
\label{sec:challenges}

To make decentralized MARL a practical paradigm for resilient critical infrastructures, two challenges are especially central: how agents identify their contributions to system-level outcomes, and how they maintain effective coordination through communication under distributed and disrupted conditions.

\subsection{Credit Assignment}

Credit assignment in MARL refers to the problem of attributing system-level outcomes such as overall performance to the actions of individual agents. While prior work often associates credit assignment with shared global rewards that cannot be decomposed into individual contributions~\cite{foerster2018coma}, the problem more fundamentally arises from the coupling between individual reward of one agent and the actions of others. This coupling may occur either through an explicit global reward or through individual rewards that are strongly influenced by the actions of others. A concrete example is decentralized traffic signal control along a congested corridor. Each intersection receives local feedback, such as whether its queue length or waiting time has decreased, but this outcome depends on the joint signal decisions of upstream and downstream intersections. When congestion persists, an individual intersection cannot easily determine whether the unresolved congestion is caused by its own phase timing, excessive inflow released by upstream intersections, or insufficient downstream capacity. This illustrates a credit assignment problem caused by strong action coupling among traffic signal agents. As a result, an intersection may reinforce ineffective local actions or overreact to congestion caused by neighboring agents. Over time, such incorrect attribution can prevent coordinated policies from emerging and reduce the network’s ability to recover from disruptions.

In centralized training, a centralized controller can simultaneously observe the states, actions, and rewards of all agents, allowing it to directly assess the contribution of each individual agent’s behavior to the rewards, and even to explicitly construct counterfactual signals that capture how the result would change if a particular agent were to act differently~\cite{foerster2018coma, kapoor2024assigning}. However, in decentralized MARL, each agent must infer how its own actions affect the rewards it receives using only local observations and limited feedback. Without access to the states and actions of other agents, it is difficult for an agent to reason whether a reward outcome is due to its own behavior or to the behavior of others, making credit assignment a central challenge in decentralized settings. Without clear ways to link individual actions to rewards, participants might focus on what benefits them locally or take advantage of others’ efforts, even when this undermines the success of the system as a whole. 

This challenge becomes especially important when decentralized MARL is applied to resilient critical infrastructures. The appeal of decentralized MARL in such settings lies in its potential to support distributed control without relying on a central coordinator. But for this potential to be realized, local agents must learn behaviors that contribute not only to local performance, but also to broader system-level resilience. In the traffic-control example, an intersection that greedily reduces its own queue may release excessive traffic downstream and cause congestion to spread to other parts of the network. Conversely, an intersection may need to temporarily accept local delay in order to support corridor-level recovery after an incident or demand surge. Credit assignment therefore matters because it shapes whether decentralized agents can learn policies that support coordinated recovery and system-level resilience rather than only isolated local improvement.

\subsection{Communication}

Communication has long been recognized as important in cooperative MARL under partial observability, especially during execution, where agents may need to exchange information to coordinate their actions effectively~\cite{zhu2024survey}. In decentralized MARL, however, the communication challenge extends beyond execution-time coordination and becomes a core part of training itself. Without a central trainer or critic to aggregate information and align learning, agents must rely on distributed communication to exchange local knowledge, coordinate policy updates, and support the emergence of effective decentralized decision-making rules. This makes communication a fundamental bottleneck for learning, rather than simply an auxiliary mechanism for action coordination. 

However, communication in decentralized environments cannot be assumed to be free or unlimited.
Because bandwidth, latency, and reliability constraints limit how much information can be exchanged and how often exchanges can occur, communication must be selective and adaptive rather than exhaustive or static. Agents must therefore determine what information is worth sharing, when it should be shared, with whom it should be shared, and how information received from others should be incorporated into learning and coordination. The challenge is further complicated by the fact that agents’ policies evolve during learning, so the meaning and usefulness of communicated information may also change over time. Communication therefore directly shapes whether distributed agents can learn coherent policies under realistic constraints.

This challenge becomes especially significant in resilient critical infrastructures, where coordination must be maintained during both normal operation and disruption recovery. In such settings, communication allows local agents to become aware of disruptions beyond their own observations and adjust their behavior as operating conditions change. When communication is impaired, agents must act on stale or partial knowledge, which can slow adaptation, allow local disruptions to propagate, and hinder recovery. Communication is therefore central to resilience in decentralized MARL because it determines whether agents can coordinate disruption responses and sustain or restore system-level performance under real-world infrastructure constraints.

\section{Research Agenda}
\label{sec:agenda}

Section \ref{sec:challenges} shows that credit assignment and communication are the two main bottlenecks that hinder the practical feasibility of decentralized MARL in resilient critical infrastructures. Building on these insights, this section translates these bottlenecks into a research agenda. The first two directions address these bottlenecks directly: credit assignment that aligns local learning with system-level resilience, and communication that supports both coordination and attribution. The third direction concerns the deployment conditions under which these mechanisms must operate, requiring decentralized learning to remain safe, timely, and recoverable under realistic infrastructure constraints.

\subsection{Structure-Aware, Causality-Aware, and Resilience-Aware Credit Assignment}

Future decentralized MARL systems for critical infrastructures should develop credit assignment mechanisms that are structure-aware, causality-aware, and resilience-aware.

Credit assignment should leverage infrastructure topology and operational dependencies to limit attribution to relevant agents and approximate their causal contributions within a localized attribution scope. When attributing an observed reward or outcome, an agent should focus on agents that are physically connected, functionally interdependent, or located along relevant propagation paths. For example, agents may construct localized counterfactual estimates by combining their own observations with compact information from relevant neighbors and estimating how the observed reward might have changed if their own action or a neighboring agent’s action had been different. This can help an agent infer whether the outcome was caused by its own behavior, by other relevant agents, or by their joint interaction.

Credit assignment should also be resilience-aware. When an agent interprets its received reward, it should consider not only immediate local efficiency, but also whether its action contributed to resilience-relevant outcomes, such as preserving essential functions, limiting degradation, or supporting recovery under disruption. In this way, credit assignment helps align local learning signals with system-level resilience objectives.

\subsection{Communication for Coordination and Credit Assignment}

Communication in decentralized MARL should be studied in two complementary roles. The first is coordination. During training, agents exchange local knowledge, policy information, value estimates, or compact summaries so that their learning updates do not evolve in isolation. During execution, agents exchange current conditions, intended actions, warnings, or local predictions to coordinate behavior under partial observability, disruption, and changing operating conditions.

The second role is to support decentralized credit assignment. Since each agent observes only part of the system, communication can provide additional evidence for assessing whether an outcome was caused by the agent itself, by nearby agents, or by their joint behavior. Such evidence may include local rewards, action intentions, local state summaries, uncertainty estimates, or compact policy information. In this sense, communication is not only a coordination channel, but also a way to improve the learning signals available to decentralized agents.

A key research direction is to design adaptive communication protocols that decide what to communicate, when to communicate, with whom to communicate, and how to use received information. Communication should be treated as a constrained resource subject to limited bandwidth, latency, reliability, and safety requirements, rather than as an always-available and unlimited auxiliary channel.

\subsection{Safe, Timely, and Recoverable Learning under Deployment Constraints}

The preceding two directions focus on how decentralized agents assign credit and communicate under limited information. For critical infrastructures, however, these mechanisms cannot be evaluated only by whether they improve learning performance or coordination efficiency. They must also remain safe, timely, and recoverable under deployment constraints. A credit assignment mechanism that encourages unsafe local actions, or a communication protocol that works only when messages arrive without delay, would not be acceptable in practice.

This requires attention to both learning design and evaluation. On the learning side, decentralized MARL should be combined with simple safety and fallback mechanisms. For example, agents should avoid actions that violate basic operational rules, exceed physical limits, or miss required decision deadlines\cite{majumdar2023discrete}. When learning or communication becomes unreliable, agents may reduce the size of policy updates, rely more on trusted local information, or temporarily switch to conservative behaviors until coordination is restored. On the evaluation side, benchmarks should test agents under representative disruptions, including communication failures, partial controller failures, abnormal operating conditions, and malicious attacks. Success should therefore not be defined only by strong performance under ideal simulation conditions, but also by whether decentralized MARL remains safe, timely, and recoverable under realistic infrastructure disruptions.

\section{Conclusion}
\label{sec:conclusion}

This paper argues that decentralized MARL is not merely a distributed alternative to centralized training with decentralized execution, but a structurally appropriate paradigm for resilient critical infrastructures. This argument is grounded in the correspondence between the properties of decentralized MARL and the operational requirements of critical infrastructure environments, including scalability, robustness, privacy, local autonomy, and interaction-driven adaptation. At the same time, this paper emphasizes that structural alignment alone is not sufficient for practical deployment. Credit assignment and communication are central conditions for realizing the potential of decentralized MARL in practice. Building on these challenges, the proposed research agenda identifies three priorities for future work: credit assignment that is structure-aware, causality-aware, and resilience-aware; communication that supports both coordination and credit assignment; and decentralized learning that remains safe, timely, and recoverable under deployment constraints. Overall, decentralized MARL should be understood as a framework for studying how resilient system-level behavior can emerge from local learning, limited information, and decentralized coordination among interdependent infrastructure components.

\section*{Acknowledgment}
This research is supported by a UKRI Future Leaders Fellowship (MR-/W009560/1): \textit{Digitally Assisted Collective Governance of Smart City Commons--ARTIO}.

\bibliographystyle{IEEEtran}
\bibliography{reference}

\end{document}